\documentclass[12pt,a4paper,oneside,final]{article}
\usepackage{amsmath}

\input{tcilatex}

\begin{document}

\title{Measures of distance in quantum mechanics}
\author{P. Gusin, D. Burys, A. Radosz \\
Department of Quantum Technologies, \\
Wroclaw 50-370, University of Science and Technology, Poland}
\maketitle

\begin{abstract}
Combining gravity with quantum theory is still work in progress. On the one
hand, classical gravity, is the geometry of space-time determined by the
energy-momentum tensor of matter and the resulting nonlinear equations; on
the other hand, the mathematical description of a quantum system, is Hilbert
space with linear equations describing evolution. In this paper, various
measures in Hilbert space will be presented. In general, distance measures
in Hilbert space can be divided into measures determined by energy and
measures determined by entropy. Entropy measures determine quasi-distance
because they do not satisfy all the axioms defining distance. Finding a
general rule to determine such a measure unambiguously seems to be
fundamental.
\end{abstract}

\section{Introduction}

What defines a metric, that is, a way to measure distances on a given
manifold? In the case of classical, i.e. non-quantum physics, the answer is
provided by the general of relativity (GR). Metrics are determined by the
distribution of masses and currents of matter. The modern approach to this
problem began with Clifford's work [1] in 1876. Thirty-nine years later,
Einstein gave the solution in the form of the general theory of relativity.

The concept of a metric is basic one and enters almost every equation of
physics. For example, in order to define one of the basic in physics
operators, the Laplace operator, it is necessary to determine a metric first.

In the case of quantum physics whose states are defined in Hilbert space,
there is no such a single measure of distance. On the contrary, there are
many measures of distance between states, some used in quantum information
theory (also applied in research related to quantum gravity, see e.g. [2])
but none of them follows from some fundamental principle. Since Hilbert
space is a vector complex space, so one can introduce a metric that is
induced from $C^{N}$\ and call this metric canonical. Giving as a result the
Fubini-Study (FS) metric. And based on it to determine the distance. In the
case when Hilbert space is represented by square-integrable functions $%
L^{2}\left( M\right) $ on some manifold $M$, then the scalar product on this
Hilbert space is given by the volume form $d\mu $\ on $M$. Again, this
scalar product leads to the FS metric. For both finite $N$\ and for $%
L^{2}\left( M\right) $, the largest distance between points of these Hilbert
spaces is normalized to $\pi $. However, such a canonical metric does not
represent the complexity of the quantum system. Nevertheless, and as is well
known, the probabilistic interpretation of the quantum system is based on
this metric.

Another issue is the geometrization of thermodynamics. There are such
metrics as Weinhold metric or Ruppeiner metric, where the relationship
between them is found. However, also in this case none of the metrics is
derived from some fundamental equation. Since widely studied and the
well-known topic is the thermodynamics of black holes, the combination of
geometrization of thermodynamics and black hole seems promising. Many papers
have been written on this topic, e.g. [3-5].

The fundamental classical concept describing a physical system is the action
integral. The energy-momentum tensor of a system results as a variation of
the action integral with respect to a metric. The general relativity links
the energy-momentum tensor to the geometric properties of space-time, i.e.,
the Riemann tensor. In contrast, the quantum concept describing the system
is the density matrix. And there is no theory or relationship that links the
density matrix to the quantum properties of space-time, because such
concepts do not exist. One may ask: are there any hints in existing theories
that would provide such ideas?

The first such a hint may be the concept of entropy determined by the
density matrix. One should look for an extension of the concept of entropy
in an analogous way to how energy density and momentum density combine in
the energy-momentum tensor.

A second such hint is geometrodynamics [6]. One way to find a quantum theory
of gravity is to try to quantize the general theory of relativity using
canonical quantization. In this approach, the Wheeler-De Witt (WDW) equation
plays a fundamental role. Many interesting results have been obtained
following this way of reasoning, e.g.[7-9]. So, the concepts that allow to
describe quantum space-time should be obtained in the framework of
geometrodynamics from the WDW equation.\textit{\ }Although this is not a
work on quantum gravity (QG), and is only an attempt to find an analogy of
the equivalence principle in quantum mechanics, we will just mention the two
most common and fruitful approaches to the issue of QM as a metric theory.
That means canonical and covariant approachs. In the 1960s, De Witt
presented attempts to quantize the gravitational field in canonical and
covariant approaches [10-12]. Since then, much has been understood and
achieved but many problems remain.\textit{\ }The actual question is whether
these approaches are equivalent or not, namely they possibly represent a
single formal theory, or not. Otherwise, the issue should be that of finding
out which of them, if any - simply on the basis of general physical
principles - may appear as the correct one. Regarding possible alternative
routes to QG theory, covariant quantization should provide hints to the
considered problem. The literature on covariant quantization of gravity is
vast, so the authors will cite only one [13] and another that is a new
approach to the Hamiltonian formulation of gravity [14-15].

Moreover, in quantum field theory in curved space-time (the zero
approximation of quantum gravity), the problem of determining the vacuum
state of the field arises. There are many vacuum states which are not
unitary related. For example in the Schwarzschild\ spacetime with free
scalar field there are three well-known vacuum states: Hartle-Hawking,
Boulware and Unruh. Each of these gives different expectation values of the
field operators [16]. Another example is de Sitter spacetime with families
of vacuua states of quantum scalar field [17]. So the natural question is
about the relationship between such states. A good measure to determine
these relationships is the distance between them.

In this paper we propose a metric in a general case of a Hilbert space of a
quantum mechanical system. This metric is infinitely dimensional version of
Fisher-Rao metric on infinitly dimensional sphere $S^{\infty }$. We apply
such a metric for simplest quantum systems: a free particle and harmonic
oscillator. The "distance" given by the relative entropy is derived and
calculated for different quantum systems.

The paper is organized as follows. In the section 2 it is recalled how the
distance is determined in classical physics. In the section 3, the metrics
in the space of probability distributions are presented and the Fisher-Rao
(FR) metric is derived as a condition for the stationarity of the "action
integral". In the section 4 we give the metric in the Hilbert space of a
quantum mechanical system. In the section 5 we find the FR metrics in the
case of a free particle and harmonic oscillator. In the section 6 we present
and calculate the "distance" given by the relative entropy. The section 7 is
devoted to conclusions.

\section{Measure of the distance in space-time}

To determine the distance in space-time $M$ with fixed symmetry and matter
with a given energy-momentum tensor $T_{\mu \nu }$, it is necessary to solve
Einstein's equations:%
\begin{equation}
G_{\mu \nu }\left( g\right) =16\pi T_{\mu \nu }\left( g\right)  \tag{2.1}
\end{equation}%
that is, to determine the metric field $g$ where $G_{\mu \nu }$ is the
Einstein tensor. This is a nontrivial task and there are few exact
solutions. From the metric "$g$" obtained in this way, it is necessary to
find the line (1-dimensional submanifold) $\gamma $ along which the distance
will be measured. The line $\gamma $ is parameterized by the affine
parameter "$s$":%
\begin{equation}
\gamma =\left\{ x\in M:\text{ }x=x\left( s\right) \right\}  \tag{2.2}
\end{equation}%
and "$x$" are coordinates on $M$. The equation of such a line, is found from
the stationarity of the functional:%
\begin{equation*}
\delta L=0,
\end{equation*}%
where:%
\begin{equation}
L\left[ \gamma \right] =\frac{1}{2}\dint\limits_{s_{1}}^{s_{2}}g_{\mu \nu
}\left( x\right) \frac{dx^{\nu }}{ds}\frac{dx^{\nu }}{ds}ds  \tag{2.3}
\end{equation}%
with the fixed boundary points: $x_{1}=x\left( s_{1}\right) $ and $%
x_{2}=x\left( s_{2}\right) $. It leads to the geodesic equation:%
\begin{equation}
\frac{d^{2}x^{\mu }}{ds^{2}}+\Gamma _{\nu \rho }^{\mu }\frac{dx^{\nu }}{ds}%
\frac{dx^{\rho }}{ds}=0,  \tag{2.4}
\end{equation}%
with the \textit{boundary} conditions $x_{1}=x\left( s_{1}\right) $ and $%
x_{2}=x\left( s_{2}\right) $. In order to solve this equation in an
unambiguous way, it is necessary to give \textit{initial} conditions:%
\begin{equation}
x\left( s_{1}\right) =x_{1}\text{ \ and }\left. \frac{dx}{ds}\right\vert
_{s=s_{1}}=\overset{\cdot }{x}_{1}.  \tag{2.5}
\end{equation}%
Thus one has to express these initial conditions by the boundary conditions.
The expression of boundary conditions by initial conditions is unambiguous
when the points $x_{1}$ and $x_{2}$ are "close". The expression "close"
means that $x_{1}$ and $x_{2}$ are not conjugate. Finally, the distance $d$
between points $x_{1}$ and $x_{2}$ is given by the integral:%
\begin{equation}
0<d\left( x_{1},x_{2}\right) =\dint\limits_{s_{1}}^{s_{2}}\sqrt{\left\vert
g_{\mu \nu }\frac{dx^{\nu }}{ds}\frac{dx^{\nu }}{ds}\right\vert }ds. 
\tag{2.6}
\end{equation}%
Since the metric $g$ at each point has a signature $(-,+,+,+)$, there are
spatial-like $d^{2}\left( x_{1},x_{2}\right) >0$, time-like $d^{2}\left(
x_{1},x_{2}\right) <0$ and light-like $d^{2}\left( x_{1},x_{2}\right) =0$
distances. So the geodesic is a line of maximum length. A line with a
minimum length of zero between two points (causally related) always exists,
and it is the sum of zero geodesics.

\section{Metrics for space of probability distributions}

For a random variable $X\in \Omega $ (where the set $\Omega $ can be
continuous or\ discrete) is given a probability distribution $p$:%
\begin{equation}
p=p\left( X;\theta _{1},...,\theta _{N}\right) ,  \tag{3.1}
\end{equation}%
where the numbers $\left( \theta _{1},...,\theta _{N}\right) \equiv \Theta $
are the parameters of the distribution $p$ belonging in the general case to
some manifold $M$. So $p$ is the function defined on the Cartesian product $%
\Omega \times M$ with values between 0 and 1 for each $\Theta $: 
\begin{equation}
p:\Omega \times M\rightarrow \lbrack 0,1].  \tag{3.2}
\end{equation}%
Moreover there is the normalization condition:%
\begin{equation}
\int_{\Omega }p\left( X;\Theta \right) d\mu =1,  \tag{3.3}
\end{equation}%
where $d\mu $ is measure on $\Omega $. This condition defines the manifold $%
M $. So the given probability distribution $p$ determines $M$ but, on the
other hand, this distribution itself depends on the points of $M$. Thus, the
space of probability distributions is obtained:%
\begin{equation}
\mathcal{P}=\left\{ p\left( X;\Theta \right) \right\}  \tag{3.4}
\end{equation}%
as the space of functions over $M$. The question arises: how can the
distance between two distributions on $\mathcal{P}$ be determined? And the
question: is there any distinguished set of metrics on $M$?

In the geometric approach, the metric is given by the relation (e.g \ [18]):%
\begin{equation}
g_{ab}\left( \Theta ;F\right) =\int_{\Omega }p\frac{\partial F}{\partial
\theta ^{a}}\frac{\partial F}{\partial \theta ^{b}}d\mu =\int_{\Omega }p%
\left[ F^{\prime }\left( p\right) \right] ^{2}\frac{\partial p}{\partial
\theta ^{a}}\frac{\partial p}{\partial \theta ^{b}}d\mu ,  \tag{3.5}
\end{equation}%
where $F$ is a function of $p$ and prime means differentiation with respect
to $p$. From this equation one can obtain an "action integral" $S$ for the
function $F$ depending on the one degree of freedom $p:$%
\begin{equation*}
S[F]=\dint L\left[ F;p\right] dp,
\end{equation*}%
where the Lagrangian $L\left[ F;p\right] $ for this action is equal to:%
\begin{equation}
L\left[ F;p\right] =p\left[ F^{\prime }\left( p\right) \right] ^{2}. 
\tag{3.6}
\end{equation}%
Hence, the stationarity condition for this "action integral" leads to the
Euler-Lagrange equation:%
\begin{equation}
\frac{d}{dp}\left( pF^{\prime }\right) =0,  \tag{3.7}
\end{equation}%
with the solution:%
\begin{equation}
F\left( p\right) =k\ln p+F_{0},  \tag{3.8}
\end{equation}%
where $k$ and $F_{0}$ are constants of integration. In this way, the metric
for this solution is of the form:%
\begin{equation}
g_{ab}\left( \Theta \right) =k^{2}\int_{\Omega }p\left( X;\Theta \right) 
\frac{\partial \ln p}{\partial \theta ^{a}}\frac{\partial \ln p}{\partial
\theta ^{b}}d\mu =k^{2}\int_{\Omega }\frac{\partial _{a}p\partial _{b}p}{p}%
d\mu .  \tag{3.9}
\end{equation}%
This is the Fisher-Rao (FS) metric. The other form of this metric is as
follows:%
\begin{equation}
g_{ab}\left( \Theta \right) =-k^{2}\int_{\Omega }p\partial _{ab}^{2}\ln
pd\mu .  \tag{3.10}
\end{equation}%
Hence, the infinitesimal square of length on $M$ has the form:%
\begin{equation}
dl^{2}=g_{ab}d\theta ^{a}d\theta ^{b}  \tag{3.11}
\end{equation}

As an example, we will consider the Gauss distribution: 
\begin{equation}
p\left( X;\theta ^{1},\theta ^{2}\right) =\frac{1}{\sqrt{2\pi }\theta ^{1}}%
\exp \left[ -\frac{\left( X-\theta ^{2}\right) ^{2}}{2\left( \theta
^{1}\right) ^{2}}\right] .  \tag{3.12}
\end{equation}%
Thus the FR metric is well-known and equal to (for $k=1$):%
\begin{equation}
dl^{2}=\frac{1}{\left( \theta ^{1}\right) ^{2}}\left( d\left( \theta
^{2}\right) ^{2}+2d\left( \theta ^{1}\right) ^{2}\right) .  \tag{3.13}
\end{equation}%
This is the metric on the Poincare upper half plane $\mathbf{H}$ defined by
the condition $\theta ^{1}>0$. Thus the geodesic distance between two Gauss
distributions:%
\begin{equation}
p_{1}=p\left( X;\Theta _{1}\right) \text{ and }p_{2}=p\left( X;\Theta
_{2}\right)  \tag{3.14}
\end{equation}%
in this metric is equal to:%
\begin{equation}
d\left( p_{1},p_{2}\right) =2\sinh ^{-1}\left[ \frac{\left\vert \Theta
_{2}-\Theta _{1}\right\vert }{2\sqrt{\theta _{2}^{1}\theta _{1}^{1}}}\right]
,  \tag{3.15}
\end{equation}%
where $\Theta _{a}=\left( \theta _{a}^{1},\theta _{a}^{2}\right) $, $a=1,2$, 
$\left\vert \Theta \right\vert =\sqrt{\left( \theta ^{1}\right) ^{2}+\left(
\theta ^{2}\right) ^{2}}$ and $\sinh ^{-1}X=\ln \left( X+\sqrt{1+X^{2}}%
\right) $. If we omit the stationarity condition, the metric for the Gauss
distribution takes the form:%
\begin{equation}
g_{11}=\frac{\sqrt{2}}{\theta ^{1}}\int_{\mathbf{R}^{1}}p^{3}\left( z\right) %
\left[ F^{\prime }\left( p\right) \right] ^{2}\left[ 1-4z^{2}+4z^{4}\right]
dz,  \tag{3.16}
\end{equation}%
\begin{equation}
g_{22}=\frac{1}{\sqrt{2}\theta ^{1}}\int_{\mathbf{R}^{1}}p^{3}\left(
z\right) \left[ F^{\prime }\left( p\right) \right] ^{2}z^{2}dz,  \tag{3.17}
\end{equation}%
\begin{equation}
g_{12}=0,  \tag{3.18}
\end{equation}%
where:%
\begin{equation}
p\left( z\right) =\frac{1}{\sqrt{2\pi }\theta ^{1}}\exp \left( -z^{2}\right)
.  \tag{3.19}
\end{equation}%
If we require that the following condition is satisfied:%
\begin{equation}
p^{3}\left[ F^{\prime }\left( p\right) \right] ^{2}=p,  \tag{3.20}
\end{equation}%
then $F\left( p\right) =\ln p$ and again the FR metric is obtained. So one
can say that for the Gaussian distribution the stationarity condition and
above condition (which can be referred to as a simplicity condition) give
the same result, namely FR metric.

\section{FR metric in quantum mechanics}

Quantum mechanics provides probability distributions $P$ expressed by wave
functions $\psi $. In general, the wave function $\Psi $ depends on time $t$%
, the parameters of the system $\omega $ and the initial state $\psi $, 
\begin{equation}
\Psi =\Psi \left( x,t,\omega ,\psi \right)  \tag{4.1}
\end{equation}%
where $x$ denote the spatial coordinates. If $\widehat{H}$ is time
independent Hamiltonian of the system, then in the eigenbasis $\psi _{n}$ of 
$\widehat{H}$ with eigenvalues $E_{n}$ the initial state $\psi
=\dsum\limits_{n=0}c_{n}\psi _{n}\left( x;\omega \right) $ evolves as
follows:%
\begin{equation}
\Psi \left( x,t,\omega ,c\right) =\exp \left( -it\widehat{H}\right) \psi
=\dsum\limits_{n}c_{n}\exp \left( -itE_{n}\left( \omega \right) \right) \psi
_{n}\left( x;\omega \right) .  \tag{4.2}
\end{equation}%
The complex coefficients $c=\left( c_{n}\right) $ are normalized: $%
\dsum\limits_{n=0}\left\vert c_{n}\right\vert ^{2}=1$. The wave function $%
\Psi $ can also be expressed via: 
\begin{equation}
\Psi \left( x;t,\omega \right) =\int_{\mathbf{R}^{N}}d^{N}yK\left(
x,t;y,0\right) \psi \left( y;\omega \right) ,  \tag{4.3}
\end{equation}%
propagator $K\left( y,0;x,t\right) $ with the initial condition:%
\begin{equation}
K\left( x,0;y,0\right) =\delta ^{\left( N\right) }\left( x-y\right) 
\tag{4.4}
\end{equation}%
and $N$ is the number degrees of freedom. In basis $\psi _{n}$ the
propagator takes the form:%
\begin{equation}
K\left( x,t;y,0\right) =\dsum\limits_{n}\exp \left( -itE_{n}\left( \omega
\right) \right) \psi _{n}^{\ast }\left( x;\omega \right) \psi _{n}\left(
y;\omega \right)  \tag{4.5}
\end{equation}%
Thus the probability distribution $P$ is equal to:%
\begin{equation}
P\left( x;t,\omega ,c\right) =\dsum\limits_{m,n}c_{m}^{\ast
}c_{n}I_{mn}\left( x,t,\omega \right) ,  \tag{4.6}
\end{equation}%
where:%
\begin{equation}
I_{mn}\left( x,t,\omega \right) =\int_{\mathbf{R}^{2N}}d^{N}yd^{N}zK^{\ast
}\left( x,t;y,0\right) K\left( x,t;z,0\right) \psi _{m}^{\ast }\left(
y;\omega \right) \psi _{n}\left( z;\omega \right) =I_{nm}^{\ast }\left(
x,t,\omega \right)  \tag{4.7}
\end{equation}%
The probability normalization condition is satisfied:%
\begin{equation}
\int_{\mathbf{R}^{N}}d^{N}xP\left( x;t,\omega ,c\right) =1.  \tag{4.8}
\end{equation}%
Thus, for a fixed time $t$ and parameters $\omega $, the FR metric is a
function of the complex numbers $c_{m}$ and the metric has the components::%
\begin{equation}
g_{mn}\left( c\right) =-\int_{\mathbf{R}^{N}}d^{N}xP\frac{\partial ^{2}\ln P%
}{\partial c_{m}\partial c_{n}},  \tag{4.9a}
\end{equation}%
\begin{equation}
g_{\overline{m}n}\left( c\right) =-\int_{\mathbf{R}^{N}}d^{N}xP\frac{%
\partial ^{2}\ln P}{\partial c_{m}^{\ast }\partial c_{n}},  \tag{4.9b}
\end{equation}%
\begin{equation}
g_{\overline{m}\overline{n}}\left( c\right) =-\int_{\mathbf{R}^{N}}d^{N}xP%
\frac{\partial ^{2}\ln P}{\partial c_{m}^{\ast }\partial c_{n}^{\ast }}%
=g_{mn}^{\ast }\left( c\right) .  \tag{4.9c}
\end{equation}%
Thus we get:%
\begin{equation}
g_{mn}\left( c;t,\omega \right) =\dsum\limits_{k,p}c_{k}^{\ast }c_{p}^{\ast
}A_{mn}^{kp}\left( c;t,\omega \right) ,  \tag{4.10a}
\end{equation}%
\begin{equation}
g_{\overline{m}n}\left( c;t,\omega \right) =-\int_{\mathbf{R}%
^{N}}d^{N}xI_{mn}\left( x,t,\omega \right)
+\dsum\limits_{k,p}c_{k}c_{p}^{\ast }A_{kn}^{mp}\left( c;t,\omega \right) , 
\tag{4.10b}
\end{equation}%
where:%
\begin{equation}
A_{mn}^{kp}\left( c;t,\omega \right) =\int_{\mathbf{R}^{N}}d^{N}x\frac{%
I_{km}\left( x,t,\omega \right) I_{pn}\left( x,t,\omega \right) }{P\left(
x;t,\omega ,c\right) }  \tag{4.11}
\end{equation}%
Since the eigenfunctions $\psi _{n}$ form a complete and orthogonal system
then the first integral on the right-hand side is equal to: 
\begin{equation}
\int_{\mathbf{R}^{N}}d^{N}xI_{mn}\left( x,t,\omega \right) =\delta _{mn}. 
\tag{4.12}
\end{equation}%
Thus:%
\begin{equation}
g_{\overline{m}n}\left( c;t,\omega \right) =-\delta
_{mn}+\dsum\limits_{k,p}c_{k}c_{p}^{\ast }A_{kn}^{mp}\left( c;t,\omega
\right) .  \tag{4.13}
\end{equation}%
The complex numbers $"c"$ form the infinitely-dimensional unit sphere $%
S^{\infty }$. Thus the obtained metric is a metric on $S^{\infty }$ and has
the form: 
\begin{equation}
ds^{2}=g_{mn}dc_{m}dc_{n}+g_{mn}^{\ast }dc_{m}^{\ast }dc_{n}^{\ast }+g_{%
\overline{m}n}dc_{m}^{\ast }dc_{n}.  \tag{4.14}
\end{equation}%
It is real since: $ds^{2}=\left( ds^{2}\right) ^{\ast }$. The distance
between two states given by two sequences $c=\left( c_{n}\right) $ and $%
c^{\prime }=\left( c_{n}^{\prime }\right) $ (which are points on $S^{\infty
} $) is given by the length of the geodesic $\gamma $ originating at $c$ and
ending at $c^{\prime }$. This geodesic is determined by the above metric.
The obtained metric (4.10a-b) is the metric on the infinitely dimensional
sphere $S^{\infty }$.

In the next section we will use the above formulas for two quantum systems.

\section{Metric for free particle and harmonic oscillator}

As the first example we consider a free quantum particle. The propagator for
the free particle (in 1D) of mass $m$ has the form:%
\begin{equation}
K\left( x,t;y,0\right) =\sqrt{\frac{m}{2\pi i\hbar t}}\exp \left[ \frac{%
im\left( x-y\right) ^{2}}{2\hbar t}\right]  \tag{5.1}
\end{equation}%
and the wave functions are indexed by the wave vector $k$:%
\begin{equation}
\psi _{k}\left( x\right) =\frac{1}{\sqrt{2\pi }}\exp \left( ikx\right) . 
\tag{5.2}
\end{equation}%
The integrals (4.7) are labeled by the wave vectors $k$ and $l$. The system
has only one parameter $m$. Thus inserting above formulas into (4.7) one
obtains:%
\begin{equation}
I_{kl}\left( x;t,m\right) =\frac{m}{\left( 2\pi \right) ^{2}\hbar t}\int_{%
\mathbf{R}^{2}}dydz\exp \left[ \frac{-im\left( x-y\right) ^{2}}{2\hbar t}+%
\frac{im\left( x-z\right) ^{2}}{2\hbar t}\right] \exp \left( ikz-ily\right) .
\tag{5.3}
\end{equation}%
This integral is easy to compute and is equal to:%
\begin{equation}
I_{kl}\left( x;t,m\right) =\frac{1}{2\pi }\exp \left[ ix\left( k-l\right) -%
\frac{i\hbar t}{2m}\left( k^{2}-l^{2}\right) \right] .  \tag{5.4}
\end{equation}%
In this way the probability distribution (4.6) is:%
\begin{equation}
P\left( x;t,m,c\right) =\frac{1}{2\pi }\dsum\limits_{k,l}c_{k}^{\ast
}c_{l}\exp \left[ ix\left( k-l\right) -\frac{i\hbar t}{2m}\left(
k^{2}-l^{2}\right) \right] ,  \tag{5.5}
\end{equation}%
where the parameters space $M$ given by condition (3.3) is infinitly
dimensional sphere parametrized by the infinite sequence $\left(
c_{n}\right) $. Thus the metric (4.10a-b) is determined by the integrals
(4.11). In the considered case they are given as follow:%
\begin{gather}
A_{ln}^{kp}\left( c;t,m\right) =\frac{1}{2\pi }\exp \left[ -\frac{i\hbar t}{%
2m}\left( k^{2}-l^{2}+p^{2}-n^{2}\right) \right] \times  \notag \\
\times \int_{\mathbf{R}^{1}}dx\frac{\exp \left[ ix\left( k-l+p-n\right) %
\right] }{\dsum\limits_{r,s}c_{r}^{\ast }c_{s}\exp \left[ ix\left(
r-s\right) -\frac{i\hbar t}{2m}\left( r^{2}-s^{2}\right) \right] }. 
\tag{5.6}
\end{gather}

As the next example we consider a one-dimensional quantum harmonic
oscillator with the energy operator:%
\begin{equation}
\widehat{H}\left( m,\omega \right) =\frac{1}{2m}\widehat{p}^{2}+\frac{1}{2}%
m\omega ^{2}\widehat{x}^{2},  \tag{5.7}
\end{equation}%
where $m$ and $\omega $ are the mass and frequency respectively. The
eigenstates $\psi _{n}$ and eigenvalues $E_{n}$ are equal to:%
\begin{equation}
\psi _{n}\left( x;m,\omega \right) =\frac{1}{\sqrt{2^{n}n!}}\left( \frac{%
\lambda ^{2}}{\pi }\right) ^{1/4}\exp \left( -\frac{\lambda ^{2}x^{2}}{2}%
\right) H_{n}\left( \lambda x\right) ,  \tag{5.8}
\end{equation}%
\begin{equation}
E_{n}=\hbar \omega \left( n+1/2\right)  \tag{5.9}
\end{equation}%
where $\lambda ^{2}=m\omega /\hbar $ and $H_{n}$ are Hermite polynomials
with $n=0,1,2...$.In this case the wave function $\Psi \left( x;t,m,\omega
\right) $ is obtained from some initial state $\Psi \left( x;0,m,\omega
\right) =\psi \left( x;m,\omega \right) :$%
\begin{equation}
\Psi \left( x;t,m,\omega \right) =\int_{\mathbf{R}^{1}}dyK\left(
y,0;x,t\right) \psi \left( y;m,\omega \right) ,  \tag{5.10}
\end{equation}%
where the propagator $K$ is equal to:%
\begin{equation}
K\left( x,t;y,0\right) =\frac{\lambda }{\sqrt{2\pi i\sin \left( \omega
t\right) }}\exp \left[ \frac{i\lambda ^{2}}{2}\left( x^{2}+y^{2}\right) \cot
\left( \omega t\right) -\frac{i\lambda ^{2}xy}{\sin \left( \omega t\right) }%
\right]  \tag{5.11}
\end{equation}%
and:%
\begin{equation*}
K\left( x,0;y,0\right) =\delta \left( x-y\right) .
\end{equation*}%
Hence the probability distribution $P$ takes the form:%
\begin{equation}
P\left( x;t,m,\omega \right) =\int_{\mathbf{R}^{2}}dydzK^{\ast }\left(
x,t;y,0\right) K\left( x,t;z,0\right) \psi ^{\ast }\left( y;m,\omega \right)
\psi \left( z;m,\omega \right) .  \tag{5.12}
\end{equation}%
Finally we get:%
\begin{equation}
P\left( x;t,m,\omega ,c_{n}\right) =\frac{\lambda ^{2}}{2\pi \sin \left(
\omega t\right) }\dsum\limits_{n=0}\left\vert c_{n}\right\vert
^{2}\left\vert I_{n}\left( x;t,m,\omega \right) \right\vert ^{2},  \tag{5.13}
\end{equation}%
where:%
\begin{gather}
I_{n}\left( x;t,m,\omega \right) =\frac{1}{\sqrt{2^{n}n!}}\left( \frac{%
\lambda ^{2}}{\pi }\right) ^{1/4}\exp \left[ -\frac{i\lambda
^{2}x^{2}e^{-i\omega t}}{2\sin \left( \omega t\right) }\right] \times  \notag
\\
\times \int_{\mathbf{R}^{1}}dyH_{n}\left( \lambda y\right) \exp \left[ \frac{%
i\lambda ^{2}e^{i\omega t}}{2\sin \left( \omega t\right) }\left(
y-xe^{-i\omega t}\right) ^{2}\right] .  \tag{5.15}
\end{gather}%
So:%
\begin{equation}
g_{mn}=-2\delta _{mn}\int_{\mathbf{R}^{1}}dx\left\vert I_{n}\left(
x;t,m,\omega \right) \right\vert ^{2}+4c_{m}c_{n}\int_{\mathbf{R}^{1}}dx%
\frac{\left\vert I_{m}\left( x;t,m,\omega \right) \right\vert ^{2}\left\vert
I_{n}\left( x;t,m,\omega \right) \right\vert ^{2}}{P\left( x;t,m,\omega
,c_{n}\right) }.  \tag{5.15}
\end{equation}%
The first integral on the right side is equal to:%
\begin{equation}
\int_{\mathbf{R}^{1}}dx\left\vert I_{n}\left( x;t,m,\omega \right)
\right\vert ^{2}=\frac{2\pi }{\lambda ^{2}}\sin \left( \omega t\right) . 
\tag{5.16}
\end{equation}%
Thus the metric is:%
\begin{equation}
g_{mn}=-\delta _{mn}\frac{4\pi }{\lambda ^{2}}\sin \left( \omega t\right)
+4c_{m}c_{n}\int_{\mathbf{R}^{1}}dx\frac{\left\vert I_{m}\left( x;t,m,\omega
\right) \right\vert ^{2}\left\vert I_{n}\left( x;t,m,\omega \right)
\right\vert ^{2}}{P\left( x;t,m,\omega ,c_{n}\right) }.  \tag{5.17}
\end{equation}

As one can see from the above examples, even in the simplest quantum systems
the determination of the Fisher-Rao metric on the infinitely dimensional
sphere is a nontrivial task. The coefficients of the metric are given by the
integrals (4.11).

However if one fixes $\left( c_{n}\right) $ on $S^{\infty }$ in the case of
oscillator (this procedure can also be applied to the free particle), the
parameter space becomes: 
\begin{equation}
M_{\left( n\right) }=\left\{ \left( m,\omega \right) :m>0\text{ and }\omega
>0\right\} \subset \mathbf{R}^{2}.  \tag{5.18}
\end{equation}%
The manifold $M_{\left( n\right) }$ is two dimensional with coordinates
given by two positive numbers $m$ and $\omega .$ This space corresponds to a
set of harmonic oscillators with different masses $m$ and frequencies $%
\omega $ being in the same state given by the sequence $\left( c_{n}\right) $%
. Hence the probability distribution related to the eigenstate $\psi _{n}$ is%
\begin{equation}
p_{n}\left( x;m,\omega \right) =\frac{1}{2^{n}n!}\frac{\lambda }{\sqrt{\pi }}%
\exp \left( -\frac{\lambda ^{2}x^{2}}{2}\right) H_{n}^{2}\left( \lambda
x\right)  \tag{5.19}
\end{equation}%
and the Fisher-Rao metric $g^{\left( n\right) }$ on $M_{\left( n\right) }$
obtained from (3.10) has the components (with $k=1$):%
\begin{equation}
g_{mm}^{\left( n\right) }=\frac{1}{2m^{2}}\left( 1-n^{2}-n\right) , 
\tag{5.20a}
\end{equation}%
\begin{equation}
g_{\omega \omega }^{\left( n\right) }=\frac{1}{2\omega ^{2}}\left(
1-n^{2}-n\right) ,  \tag{5.20b}
\end{equation}%
\begin{equation}
g_{m\omega }^{\left( n\right) }=\frac{1}{4m\omega }\left( 1-n^{2}+3n\right) .
\tag{5.20c}
\end{equation}%
The line element $dl_{\left( n\right) }^{2}$ is equal to:%
\begin{equation}
dl_{\left( n\right) }^{2}=a\frac{dm^{2}}{m^{2}}+a\frac{d\omega ^{2}}{\omega
^{2}}+b\frac{dmd\omega }{m\omega },  \tag{5.21}
\end{equation}%
where $a=\left( 1-n^{2}-n\right) /2$, $b=(1-n^{2}+3n)/2$. In the coordinates 
$U$ and $V$ defined as follows: 
\begin{equation}
U=\ln \left[ \frac{\omega }{\omega _{0}}\left( \frac{m}{m_{0}}\right)
^{b/\left( 2a\right) }\right] ,  \tag{5.22a}
\end{equation}%
\begin{equation}
V=\ln \frac{m}{m_{0}},  \tag{5.22b}
\end{equation}%
the metric (5.21) takes the diagonal form: 
\begin{equation}
dl_{\left( n\right) }^{2}=adU^{2}+\eta \left( n\right) dV^{2}  \tag{5.23}
\end{equation}%
with%
\begin{equation}
\eta \left( n\right) =\frac{1}{8}\frac{\left( 1-n^{2}-5n\right) \left(
3-3n^{2}+n\right) }{1-n^{2}-n}.  \tag{5.24}
\end{equation}%
The constants $m_{0}$ and $\omega _{0}$ have the mass and $s^{-1}$
dimensions respectively and have to be determined. Since $m_{0}\omega
_{0}/\hbar $ has dimension $\left( length\right) ^{-2}$ we make following
ansatz:%
\begin{equation*}
\lambda _{0}^{2}=\frac{m_{0}\omega _{0}}{\hbar }=l_{Pl}^{-2}=\frac{c^{3}}{%
\hbar G},
\end{equation*}%
where $l_{Pl}$ is Planck length, so:%
\begin{equation*}
\omega _{0}m_{0}=\frac{c^{3}}{G}.
\end{equation*}%
For the\ successive $n$, the metric takes the form: 
\begin{equation*}
dl_{\left( 0\right) }^{2}=\frac{1}{2}dU^{2}+\frac{3}{8}dV^{2},
\end{equation*}%
\begin{equation*}
dl_{\left( 1\right) }^{2}=-\frac{1}{2}dU^{2}+\frac{5}{8}dV^{2},
\end{equation*}%
\begin{equation*}
dl_{\left( 2\right) }^{2}=-\frac{5}{2}dU^{2}-\frac{13}{8}dV^{2}.
\end{equation*}%
It can be seen from this that for $n\geq 2$ the distance becomes purely
imaginary.

\section{Entropy as the measure of a distance in Hilbert space}

A measure of a "distance" between states $\widehat{\rho }$ and $\widehat{%
\sigma }$ is the relative entropy $S_{rel}$ :%
\begin{equation}
S_{rel}\left( \widehat{\rho }||\widehat{\sigma }\right) =Tr\left[ \widehat{%
\rho }\log \widehat{\rho }-\widehat{\rho }\log \widehat{\sigma }\right] . 
\tag{6.1}
\end{equation}%
It can be rewritten as follows%
\begin{equation}
S_{rel}\left( \widehat{\rho }||\widehat{\sigma }\right) =S\left[ \widehat{%
\sigma }\right] -S\left[ \widehat{\rho }\right] +Tr\left[ \left( \widehat{%
\sigma }-\widehat{\rho }\right) \log \widehat{\sigma }\right] ,  \tag{6.2}
\end{equation}%
where:%
\begin{equation*}
S\left[ \widehat{\sigma }\right] =-Tr\left( \widehat{\sigma }\log \widehat{%
\sigma }\right)
\end{equation*}%
is von Neumana entropy. The entropy $S_{rel}$ is positive : $S_{rel}\left( 
\widehat{\rho }||\widehat{\sigma }\right) \geq 0$ and is equal to zero if $%
\widehat{\rho }=\widehat{\sigma }$. Thus one can consider this entropy as
the as the square of the "distance" between states [18, p.44]. We compute
this distance for pure states. Let the pure states have the form: $\widehat{%
\sigma }=diag(0,..,1,0,..,0)$ ($1$ is in $a-th$ place) and $\widehat{\rho }%
=diag(0,..,1,0,..,0)$ ($1$ is in $b-th$ place). Since that $\left( I-%
\widehat{\sigma }\right) ^{n}=I-\widehat{\sigma }$ and 
\begin{equation*}
\ln \left( \widehat{\sigma }\right) =\ln \left[ I-\left( I-\widehat{\sigma }%
\right) \right] =-\sum_{n=1}\frac{\left( I-\widehat{\sigma }\right) ^{n}}{n}
\end{equation*}%
one obtains that $S_{rel}\left( \widehat{\rho }||\widehat{\sigma }\right) $
is equal to:%
\begin{equation}
S_{rel}\left( \widehat{\rho }||\widehat{\sigma }\right) =-Tr\left[ \left( 
\widehat{\sigma }-\widehat{\rho }\right) \left( I-\widehat{\sigma }\right) %
\right] \sum_{n=1}\frac{1}{n}.  \tag{6.3}
\end{equation}%
Using relations: $\widehat{\sigma }=\widehat{\sigma }^{2}$ and $Tr\left( 
\widehat{\sigma }\widehat{\rho }\right) =0$ finally we get:%
\begin{equation}
S_{rel}\left( \widehat{\rho }||\widehat{\sigma }\right) =Tr\left[ \widehat{%
\rho }\right] \sum_{n=1}\frac{1}{n}=+\infty .  \tag{6.4}
\end{equation}%
Hence the square of the "distance" between two pure states is infinite! In
the case when $\widehat{\sigma }$ is pure state and $\widehat{\rho }$ is
mixed state then:%
\begin{gather}
S_{rel}\left( \widehat{\rho }||\widehat{\sigma }\right) =-S\left[ \widehat{%
\rho }\right] -Tr\left[ \widehat{\rho }\log \widehat{\sigma }\right] = 
\notag \\
\left( 1-Tr\widehat{\rho }\widehat{\sigma }\right) \sum_{n=1}\frac{1}{n}-S%
\left[ \widehat{\rho }\right] .  \tag{6.5}
\end{gather}%
However in this case this square of the "distance" can be finite, because
the difference of the infinity sum of the harmonic series and entropy of the
mixed state (which can also be infinity) could give finite results.

\subsection{Thermal states and relative entropy}

The thermal state is defined by maximum entropy $S$ and fixed energy $E$.
Thus for an energy operator $\widehat{H}$ a density matrix $\widehat{\sigma }
$ is equal to:%
\begin{equation}
\widehat{\sigma }_{t}\left( \beta \right) =\frac{1}{Z}\exp \left( -\beta 
\widehat{H}\right) ,  \tag{6.6}
\end{equation}%
where $Z=Tr\left[ \exp \left( -\beta \widehat{H}\right) \right] $ and $\beta
=1/\left( k_{B}T\right) $ is the Lagrange multiplier, which has
interpretations of the inverse of temperature $T$. The relative entropy $%
S_{rel}$ between the states $\widehat{\sigma }_{t}$ and $\widehat{\rho }$
takes the form:%
\begin{equation}
S_{rel}\left( \widehat{\rho }||\widehat{\sigma }_{t}\right) =\beta Tr\left( 
\widehat{\rho }\widehat{H}\right) -\beta Tr\left( \widehat{\sigma }_{t}%
\widehat{H}\right) -\left( S\left[ \widehat{\rho }\right] -S\left[ \widehat{%
\sigma }_{t};\beta \right] \right) ,  \tag{6.7}
\end{equation}%
where $S\left[ \widehat{\sigma }_{t};\beta \right] $ is maximum entropy $S$
and the term $Tr\left( \widehat{\sigma }\widehat{_{t}H}\right) $ is fixed
energy $E$:%
\begin{equation}
Tr\left( \widehat{\sigma }_{t}\widehat{H}\right) =E\left( \beta \right) . 
\tag{6.8}
\end{equation}%
Hence:%
\begin{equation}
S_{rel}\left( \widehat{\rho }||\widehat{\sigma }_{t}\right) =\beta Tr\left( 
\widehat{\rho }\widehat{H}\right) -S\left[ \widehat{\rho }\right] +S\left[ 
\widehat{\sigma }_{t};\beta \right] -\beta E\left( \beta \right) .  \tag{6.9}
\end{equation}%
The last two terms are related to free energy: $F\left( \beta \right)
=E-S/\beta $. Finally the relative entropy is equal to:%
\begin{equation}
S_{rel}\left( \widehat{\rho }||\widehat{\sigma }_{t}\right) =\beta Tr\left( 
\widehat{\rho }\widehat{H}\right) -S\left[ \widehat{\rho }\right] -\beta
F\left( \beta \right) .  \tag{6.10}
\end{equation}

If also $\widehat{\rho }$ \ is thermal state with an energy operator $%
\widehat{h}$%
\begin{equation}
\widehat{\rho }_{t}\left( b\right) =\frac{1}{U}\exp \left( -b\widehat{h}%
\right) ,  \tag{6.11}
\end{equation}%
[where $U=Tr\exp \left( -b\widehat{h}\right) $], then their relative entropy
is:%
\begin{equation}
S_{rel}\left( \widehat{\rho }_{t}||\widehat{\sigma }\right) =-S\left[ 
\widehat{\rho }_{t};b\right] -\frac{1}{U}Tr\left[ \exp \left( -b\widehat{h}%
\right) \log \widehat{\sigma }\right] .  \tag{6.12}
\end{equation}%
The other form of the last equation is:%
\begin{equation}
S_{rel}\left( \widehat{\rho }_{t}||\widehat{\sigma }_{t}\right) =S\left[ 
\widehat{\sigma }_{t};\beta \right] -S\left[ \widehat{\rho }_{t};b\right]
-\beta Tr\left( \widehat{\sigma }\widehat{H}\right) +\frac{\beta }{U}Tr\left[
\exp \left( -b\widehat{h}\right) \widehat{H}\right] .  \tag{6.13}
\end{equation}%
In this way the relative entropy for two thermal states with the fixed
energy operators $\widehat{H}$ and $\widehat{h}$ is parametrized by the two
positive numbers $\beta $ and $b$, with an interpretation of the inverse
temperatures (modulo Boltzman constant).

When $\widehat{H}=\widehat{h}$, then the last relation becomes:%
\begin{equation}
S_{rel}\left( \widehat{\rho }_{t}||\widehat{\sigma }_{t}\right) =S\left[ 
\widehat{\sigma }_{t};\beta \right] -S\left[ \widehat{\rho }_{t};b\right]
+\beta E\left( b\right) -\beta E\left( \beta \right) .  \tag{6.14}
\end{equation}

As an example, we will find the relative entropy for a free scalar field in
thermodynamic equilibrium. Such a field is equivalent to an infinite set of
non-interacting harmonic oscillators with maximum entropy\ $S$ and fixed
energy $E$. The entropy is equal to:%
\begin{equation}
S\left[ \widehat{\sigma }_{t};\beta \right] =2^{4}\frac{\pi ^{5}k_{B}}{45}%
\frac{1}{\left( \hbar c\right) ^{3}}\frac{V}{\beta ^{3}},  \tag{6.15}
\end{equation}%
and the energy $E\left( \beta \right) $ is equal to:%
\begin{equation}
E\left( \beta \right) =\frac{4\pi \cdot 3!}{\left( \hbar c\right) ^{3}\beta
^{4}}V\frac{\pi ^{4}}{90}+E_{0}  \tag{6.16}
\end{equation}%
where $E_{0}$ is the infinite ground-state energy and $V$ is volume of
space. Hence the relative entropy takes the form:%
\begin{equation}
S_{rel}\left( \widehat{\rho }_{t}||\widehat{\sigma }_{t}\right) =\frac{4\pi
^{5}Vk_{B}}{45\left( \hbar c\right) ^{3}\beta ^{3}}\left[ 1-4\left( \frac{%
\beta }{b}\right) ^{3}+3\left( \frac{\beta }{b}\right) ^{4}\right] . 
\tag{6.17}
\end{equation}%
If the parameter $\beta $ is in the vicinity of $b$:%
\begin{equation}
\beta =b+\delta  \tag{6.18}
\end{equation}%
for $0<\delta /b<<1$, then their relative entropy is equal to:%
\begin{equation}
S_{rel}\left( \widehat{\rho }_{t}\left( \delta \right) ||\widehat{\rho }%
_{t}\left( b+\delta \right) \right) =\frac{4\pi ^{5}Vk_{B}}{45\left( \hbar
c\right) ^{3}}6\frac{\delta ^{2}}{b^{5}}+O\left( \delta ^{3}\right) . 
\tag{6.19}
\end{equation}%
Taking the interpretation of relative entropy as the square of the distance
between states, the above relationship can be written as an expression for a
one-dimensional (one parameter\ $b$) metric:%
\begin{equation}
dl^{2}=A\frac{db^{2}}{b^{5}},  \tag{6.20}
\end{equation}%
where $A=8\pi ^{5}Vk_{B}/\left[ 15\left( \hbar c\right) ^{3}\right] $. In
term of the energy $E$ the last formula takes the form:%
\begin{equation}
dl^{2}=\frac{k_{b}}{8}\left[ \frac{4\pi ^{5}V}{15\left( \hbar c\right) ^{3}}%
\right] ^{1/4}E^{-5/4}dE^{2}.  \tag{6.21}
\end{equation}%
So the distance in this metric between two states of a free scalar field
with fixed energies $E_{2}>E_{1}$ is equal to:%
\begin{equation}
d\left( E_{1},E_{0}\right) =\frac{1}{3}\sqrt{8k_{B}}\left[ \frac{4\pi ^{5}V}{%
15\left( \hbar c\right) ^{3}}\right] ^{1/8}\left(
E_{2}^{3/8}-E_{1}^{3/8}\right) .  \tag{6.22}
\end{equation}

\section{Conclusion}

The method of determining metric and distance in classical and quantum
physics is shown in the diagram below. The left column refers to classical
physics and the right column refers to quantum physics: 
\begin{equation*}
\begin{array}{ccc}
\left( E,\mathbf{p}\right) &  & \left( \widehat{\rho },?\right) \\ 
\downarrow &  & \downarrow \\ 
\left( T_{\mu \nu }\right) &  & 
\begin{array}{c}
\left( S,?\right) \\ 
\downarrow%
\end{array}
\\ 
\begin{array}{c}
\begin{array}{c}
\downarrow \\ 
\left( \left( T_{\mu \nu }\right) ,\left( R_{\nu \rho \sigma }^{\mu }\right)
\right) \\ 
\downarrow (\text{EP)}%
\end{array}
\\ 
\delta \int \left( R+\Lambda +L_{m}\right) =0 \\ 
\downarrow \\ 
d=d\left( x,y;\left( g_{\mu \nu }\right) \right)%
\end{array}
& 
\begin{array}{c}
\Longleftarrow \\ 
?%
\end{array}
& 
\begin{array}{c}
\mathcal{H} \\ 
\downarrow \\ 
\widehat{H}\psi =i\partial _{t}\psi \\ 
\downarrow \\ 
d\left( \widehat{\rho },\widehat{\sigma }\right) =...%
\end{array}%
\end{array}%
,
\end{equation*}%
where (EP) means equivalence principle. As it follows, EP is the key idea
that is needed to determine the geometry of spacetime. Question marks
indicate unknown complementary concepts. It can be said that an ambiguity in
determining the distance or metric in quantum physics (e.g [19-21]) is due
to the lack of an counterpart of the equivalence principle.

Moreover there is an intriguing relationship between pure states and the
Kasner metric in $n+1$ dimensional spacetime. This metric describes vacuum
cosmological solutions and has the form:%
\begin{equation*}
ds^{2}=dt^{2}-\dsum\limits_{a=1}^{n}t^{2p_{a}}dx_{a}^{2},
\end{equation*}%
where the numbers $p_{1},...,p_{n}$ obey two constraints:%
\begin{equation*}
\dsum\limits_{a=1}^{n}p_{a}=\dsum\limits_{a=1}^{n}p_{a}^{2}=1.
\end{equation*}%
Thus, formally, one can consider these numbers $\left\{ p_{a}\right\} $ as
the eigenvalues of a certain density matrix $\widehat{\rho }$ corresponding
to pure state.

The obtained metric (4.10a-b) on infinitly dimensional sphere $S^{\infty }$
is given by the integrals (4.11) and was obtained as the FR metric for the
infinitly dimensional Hilbert space. It is known that when the Hilbert space
is $CP^{N}$, then the FR metric becomes the Fubini-Study metric. Thus the
metric (4.10a-b) should be reduced to Fubini-Study in the case of finite
dimensions. But it is not obvious. The sphere $S^{\infty }$ should be
modeled by some kind of limit of N-dimensional spheres $S^{N}:$%
\begin{equation*}
"\lim_{N\rightarrow \infty }"S^{N}=S^{\infty }.
\end{equation*}%
Moreover the metrics on $S^{N}$ have to be invariant under the unitary grup $%
U\left( N\right) $. For finite $N$ the normalization condtion leads to odd
value of $N=2n+1$ so the sphere $S^{2n+1}$ is represented as the homogenous
manifold:%
\begin{equation*}
\frac{U\left( n+1\right) }{U\left( n\right) }\simeq S^{2n+1}.
\end{equation*}%
From the other side $CP^{n}$ is represented also as the homogenous manifold:%
\begin{equation*}
\frac{S^{2n+1}}{S^{1}}\simeq CP^{n}.
\end{equation*}%
One of the $U\left( n\right) $ invariant metric is Fubini-Study (FS) metric.
But there are other metrics which remain hidden. In the FS metric the volume
of $CP^{N}$ is equal to:%
\begin{equation*}
vol_{FS}\left( CP^{N}\right) =\frac{\pi ^{N}}{N!}.
\end{equation*}%
Hence for $N\rightarrow \infty $ the volume in the FS metric tends to zero.
It can be seen from here that FS metric do not lead to the metric (4.10a-b).

In the case of a measure related to relative entropy, completely different
distances are obtained [e.g.21-25] , which should not be surprising. Without
some fundamental principle each measure is equally appropriate. In this
article, we have presented only very specific measures. Of course, there are
more of them.

One of the unsolved problems in quantum cosmology is finding the "norm" of
the universe's wave function. In this case, the WDW equation is defined on a
minisuperspace and the wave function is known for different boundary
conditions [26-33]. So the use of the approach outlined in this work is most
reasonable. Finding the "distance" for the wave functions describing the
expanding universe and the collapsing universe is a task to be accomplished.
A more difficult issue is the "distance" between states associated with
black holes. These issues will be addressed in the next paper.

\bigskip

Author Contributions: Conceptualization, PG.; methodology, PG, DB, AR. All
authors have read and agreed to the published version of the manuscript.

Funding: This research received no external funding.

Data Availability Statement: Data are contained within the article.

Conflicts of Interest: The authors declare no conflicts of interest.

\section{References}

[1] W. K. Clifford, \textit{On the Space Theory of Matte}r, Proc. Cambridge.
Phil. Soc. 2 (1864 -- 1876) 157 -- 158

[2] L. Susskind, \textit{Three Lectures on Complexity and Black Holes},
[arXiv:1810.11563],

[3] J. Aman, I. Bengtsson, N. Pidokrajt, \textit{Geometry of black hole
thermodynamics}, Gen.Rel.Grav.35 (2003)1733, [arXiv:gr-qc/0304015];

[4] H. Quevedo, \textit{Geometrothermodynamics}, J.Math.Phys. 48 (2007)
013506, [arXiv:physics/0604164];

[5]S. Soroushfar, R. Saffari, S. Upadhyay, \textit{Thermodynamic geometry of
a black hole surrounded by perfect fluid in Rastall theory}, Gen. Rel. Grav.
51 (2019) 130, [arXiv:1908.02133].

[6] C. W. Misner, K. S Thorne, J. A. Wheeler, \textit{Gravitation}; Freeman:
San Francisco, 1973

[7] R. Graham, P. Sz\'{e}pfalusy, \textit{Quantum creation of a generic
universe}, Phys. Rev. D 42, 2483-2490, (1990);

[8] P. Gusin, \textit{Wheeler-DeWitt equation for brane gravity}, Phys. Rev.
D 77, 066017 (2008);

[9] H. S. Vieira and V. B. Bezerra, \textit{Class of solutions of the
Wheeler-DeWitt equation in the Friedmann-Robertson-Walker universe}, Phys.
Rev. D 94, 023511 (2016), [arXiv:1603.02236]

[10] B. S. DeWitt, \textit{Quantum Theory of Gravity. 1}. \textit{The
Canonical Theory}, Phys. Rev. 160, 1113 (1967).

[11] B. S. DeWitt, \textit{Quantum Theory of Gravity. 2. The Manifestly
Covariant Theory}, Phys. Rev. 162 (1967) 1195.

[12] B. S. DeWitt, \textit{Quantum Theory of Gravity. 3. Applications of the
Covariant Theory}, Phys. Rev. 162 (1967) 1239

[13] R. Percacci, \textit{An Introduction to Covariant Quantum Gravity and
Asymptotic Safety}, World Scientific, 2017

[14] C. Cremaschini, M. Tessarotto,\ \textit{Hamiltonian approach to GR --
Part 1: covariant theory of classical gravity}, Eur. Phys. J. C (2017) 77:
329;

[15] C. Cremaschini, M. Tessarotto, \textit{Hamiltonian approach to GR --
Part 2: covariant theory of quantum gravity}, Eur. Phys. J. C (2017) 77: 330.

[16] P. Candelas, \textit{Vacuum polarization in Schwarzschild spacetime},
Phys. Rev. D 21, 2185 (1980).

[17] B. Allen, \textit{Vacuum States in de Sitter Space}, Phys. Rev. D 32,
3136 (1985).

[18] I. Bengtsson, K. \.{Z}yczkowski, \textit{Geometry of Quantum States, An
Introduction to Quantum Entanglement}, Cambridge University Press 2017.

[19] P. Facchi, R. Kulkarni, V. I. Man'ko, G. Marmo, E. C. G. Sudarshan, F.
Ventriglia, \textit{Classical and Quantum Fisher Information in the
Geometrical Formulation of Quantum Mechanics}, Physics Letters A 374 (2010)
4801, [arXiv:1009.5219].

[20] D. Bures, \textit{An extension of Kakutani's theorem on infinite
product measures to the tensor product of semifinite }$w^{\ast }$\textit{%
-algebras}, Trans. Amer. Math. Soc. 135 (1969). 199-212

[21] J. Dittman, \textit{Note on Explicit Formulae for the Bures Metric},
J.Phys.A32:2663-2670,(1999), [arXiv:quant-ph/9808044]

[22] M. A. Nielsen, I. L. Chuang, \textit{Quantum Computation and Quantum
Information}, Cambridge 2010

[23] I. D'Amico, J. P. Coe, V. V. Franca, K. Capelle, \textit{Quantum
mechanics in metric space: wave functions and their densities}, Phys. Rev.
Lett. 106, 050401 (2011) [arXiv:1102.2329]

[24] Ch. J. Turner, K. Meichanetzidis, Z. Papic, J. K. Pachos, \textit{%
Optimal free descriptions of many-body theories}, Nature Communications 8,
14926 (2017), [arXiv:1607.02679].

[25] F. M. Ciaglia, F. Di Cosmo, D. Felice, S. Mancini, G. Marmo, J. M. P%
\'{e}rez-Pardo, \textit{Aspects of geodesical motion with Fisher-Rao metric:
classical and quantum}, Open Systems \& Information Dynamics,Vol. 25, No. 1
(2018) 1850005 [arXiv:1608.06105]

[26] J. B Hartle, S. W Hawking, \textit{Wave function of the universe},
Phys. Rev. D, 28(12):2960, (1983).

[27] S. W Hawking. \textit{The quantum state of the universe}. Nuclear
Physics B, 239(1):257, (1984)

[28] Don N. Page. \textit{Hawking's wave function for the universe,}. In R.
Penrose \& C. J. Isham (eds.), \textit{Quantum Concepts in Space and Time}
(1986).

[29] A. Vilenkin, \textit{Boundary conditions in quantum cosmology,} Phys.l
Rev. D, 33(12):3560, (1986)

[30] A. Lukas. \textit{The no-boundary wave function and the duration of the
inflationary period}, Phys. Letters B, 347(1-2):13, (1995)

[31] J. B Hartle, S. W Hawking,T. Hertog, \textit{No-boundary measure of the
universe}, Phys. Rev. Letters, 100(20):201301, (2008)

[32] J. D. Dorronsoro, J. J. Halliwell, J. B. Hartle, T. Hertog, O. Janssen, 
\textit{Real no-boundary wave function in Lorentzian quantum cosmology},
Phys. Rev. D96 no. 4, (2017) 043505, [arXiv:1705.05340].

[33] J. J Halliwell, J. B Hartle,. T. Hertog, \textit{What is the
no-boundary wave function of the universe?} Phys. Rev. D, 99(4):043526,
(2019), [arXiv:1812.01760].

\bigskip

\end{document}